\documentclass[conference,usenatbib]{basi}
\begin{document}
\title[Is the `disappearance' of low-frequency QPOs in the power spectra ...]{Is the `disappearance'
of low-frequency QPOs in the power spectra a general phenomenon for Disk-Jet symbiosis?}
\author[A.~Nandi et~al.]%
       {A.~Nandi$^1$\thanks{email: \texttt{anuj@isac.gov.in}},
       D.~Radhika$^{1,2}$ and S.~Seetha$^3$\\
       $^1$Space Astronomy Group, SSIF/ISITE Campus, ISRO Satellite Centre, Bangalore, India\\
       $^2$Department of Physics, University of Calicut, Calicut, India\\
       $^3$Space Science Office, ISRO Headquarters, Bangalore, India}

\pubyear{2013}
\volume{**}
\pagerange{**--**}

\date{Received --- ; accepted ---}

\maketitle
\label{firstpage}

\begin{abstract}
One of the best possible ways to look for disk-Jet symbiosis in galactic Black Holes is to study the 
correlation between X-ray and radio emissions. Beyond this study, is there any 
alternative way to 
trace the symbiosis? To answer, we investigated the X-ray features of few black hole candidates based 
on the archival data of PCA/RXTE. We found evidences of `disappearance' of QPOs in the power density 
spectra and subsequent spectral softening of the energy spectra during the radio flares (i.e., 
`transient' Jets). We delve deep into the nature of the accretion dynamics to understand the disk-Jet 
symbiosis.  
\end{abstract}

\begin{keywords}
Black Holes, X-ray sources,  Accretion Physics, Radiation hydrodynamics
\end{keywords}

\section{Introduction}\label{s:intro}

Most of the Galactic Black Hole (GBH) sources are observed to be outbursting in nature and these
sources also have Jet emissions, which are observed as Radio flares. It is inferred that during
Jet ejections, there is an absence of QPO which implies that the inner part of the disk (i.e., `hot'
corona) gets disrupted and evacuated, and subsequently the source spectra softens implying the X-ray
emission to be mostly from the disk \citep{FM01, VS01, AN01, FP09, MJ12}.

\section{Observation and Analysis}\label{s:ObsnAna}

We analysed the public archival data obtained from HEASARC database of the RXTE Satellite to study
the temporal and spectral evolution of the BH sources during the radio flares. The standard 
procedure for PCA and HEXTE data reduction was employed using the FTOOLS package HEASOFT v 6.11. For 
timing analysis, we used PCA science data of Binned mode and Event mode. Energy dependent study of 
the power density spectra (PDS) as well as the phase lags were performed using GHATS 
v 1.0.1\footnote{http://www.brera.inaf.it/utenti/belloni/GHATS\_Package/Home.html}. 
Spectral data were extracted using Standard2 data product in the energy range of 3 - 20 keV (using 
PCU2 only). 
High energy spectral data of 20 - 150 keV were also extracted using the HEXTE data for whichever 
cluster (A/B) carried out the observations. Broadband spectral modelling was done in the energy
range of 3 - 150 keV using a thermal {\it diskbb} component, a non-thermal component
{\it powerlaw/highecut} modified by the interstellar absorption {\it phabs}.

%
%

\section{Results}

\subsection{XTE J1859$+$226}\label{s:src1}
\begin{figure}
\centerline{\includegraphics[width=8cm]{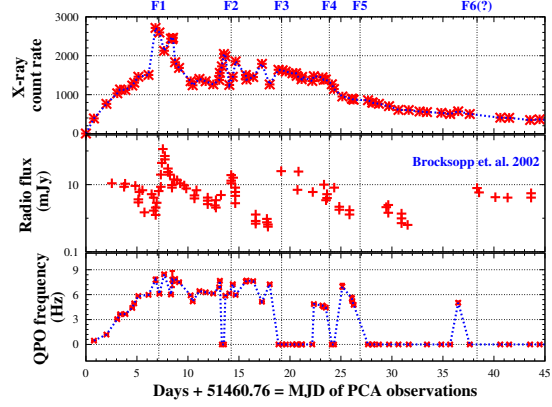}}
\caption{Variation of X-ray flux, Radio flux and QPO frequency for the BH source XTE J1859$+$226 
(1999 outburst). \label{f:one}}
\end{figure}

During the 1999 outburst of XTE J1859$+$226 (top panel of Figure \ref{f:one}), multiple Radio flares 
(see \citealt{Brock2002,FP09}) have been detected (middle panel of Figure \ref{f:one}). From the spectral 
and temporal properties of the source during these flares, we observed that during the first flare (F1), 
there is a partial `disappearance' of QPO in 2 - 5 keV and 13 - 25 keV energy bands. During the flares F2, 
F3, F4 and F5, we observed a complete absence of QPO in power spectra over 2 - 25 keV energy 
band (see \citealt{RN13} for details). We also observed that during all these flares when the 
QPO is absent, the thermal flux increases compared to the hard X-ray flux and the spectra gets soften. 
It was observed that during the flare the total rms of the PDS reduces and phase lag 
studies showed that there was no lag 
observed between the soft (2 - 6 keV) and hard (6 - 25 keV) energy bands. `Spectro-temporal' signatures 
suggest the possible presence of another flare F6, which was probably not reported due to lack of 
continuous radio observations. We observed that during F1, F2, F3, when the QPO was observed before the 
flares, it was of type C/C* (see \citealt{Wiji99,Casella2004} for details on types of 
QPOs) and when it `re-appeared' after the flare it was of type B. We observed a type B/C* QPO 
before the flare and type C* QPOs after the flare, respectively for the flares F4/F5 (\citealt{RN13}). 

\subsection{XTE J1748$-$288}\label{s:src2}
\begin{figure}
\centerline{\includegraphics[width=6cm]{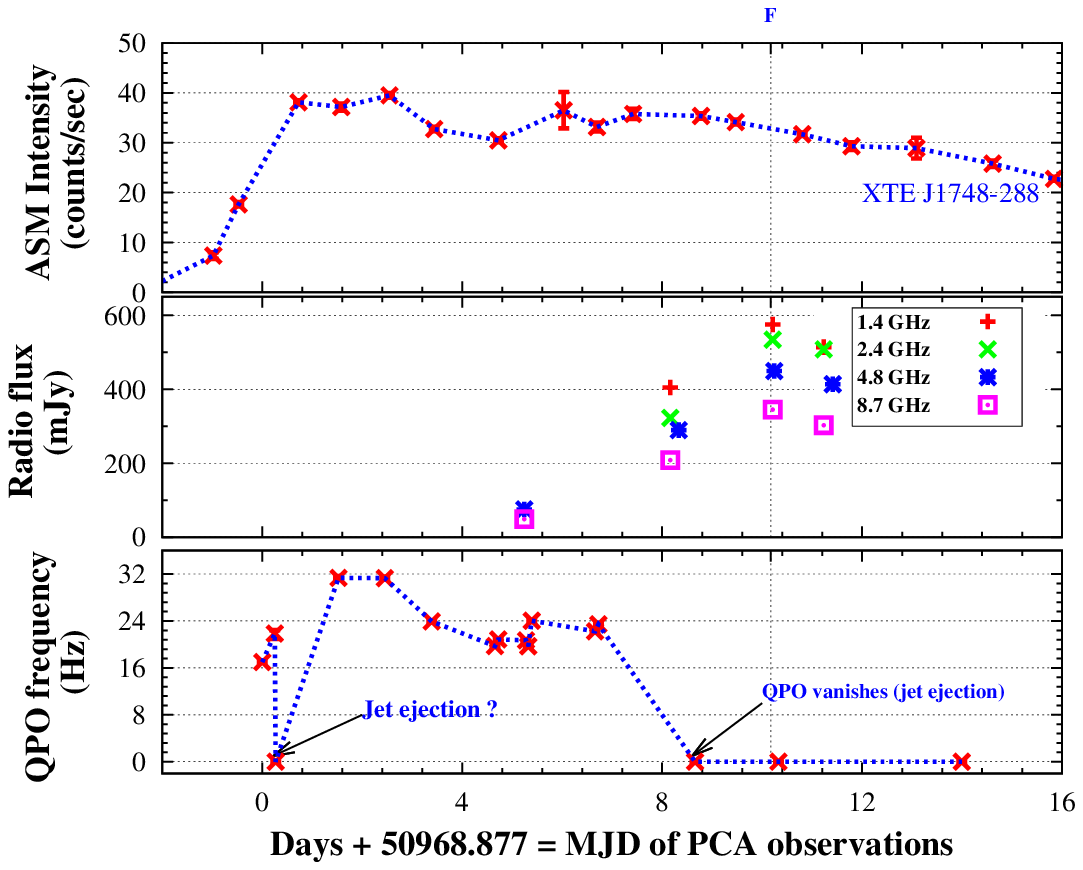} \qquad
            \includegraphics[width=6cm]{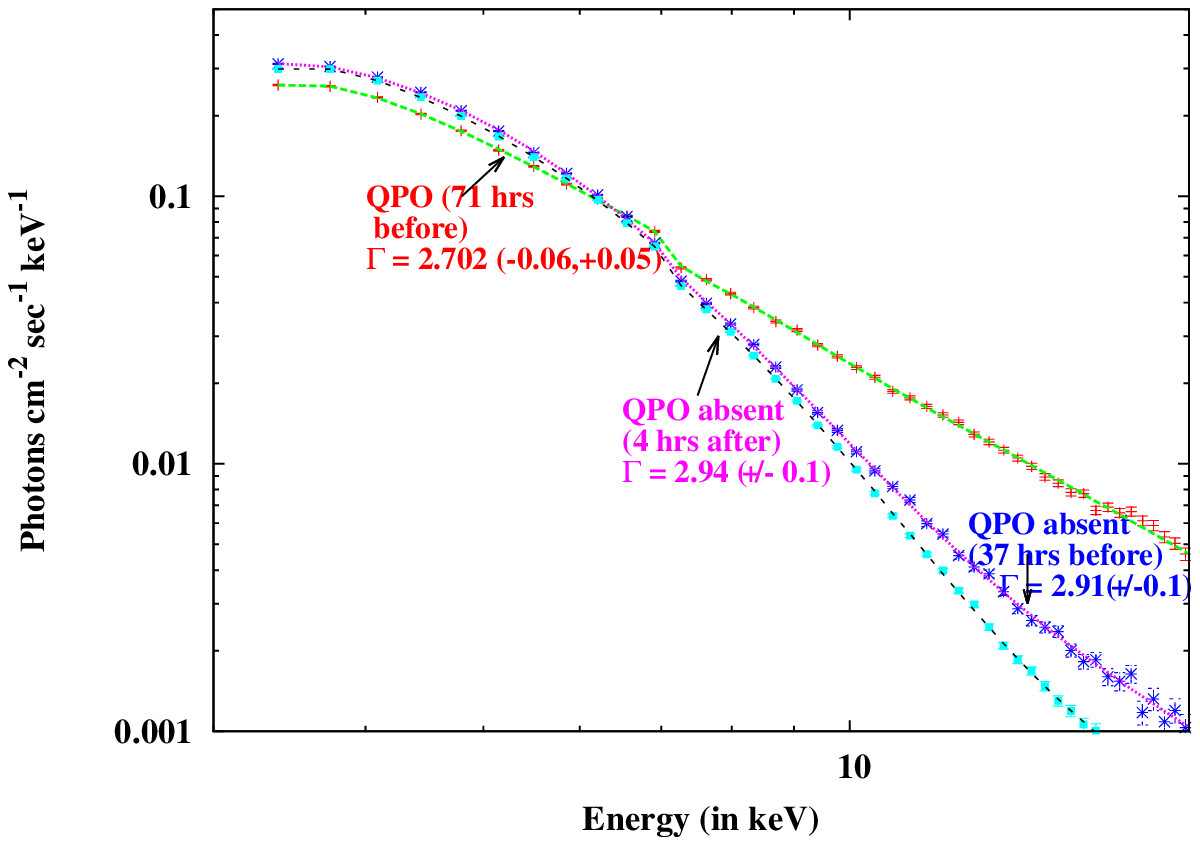}}

\caption{Left side plot shows variation of X-ray intensity, Radio flux and QPO frequency observed for the 
BH source XTE J1748$-$288 (1998 outburst). Right side plot shows spectral softening during the radio flare. 
\label{f:2}}
\end{figure}

For the 1998 outburst of XTE J1748$-$288, we observed complete absence of QPO around 24 hrs before the 
peak flare (marked as F) of 600 mJy at 1.4 GHz (see left side of Figure \ref{f:2}). We also 
observed decrease in total rms of the PDS, absence of phase lag and subsequent increase in soft 
flux which was 
also implied by the steepening of the spectral index (see right side of Figure \ref{f:2}), during the 
flare. The QPOs observed before the ejection event was found to be of type C. Absence of QPO 
and spectral softening (early phase of the outburst, when source transit from hard to soft-intermediate 
state), suggests presence of another possible ejection, although there was no radio observation 
(see \citealt{RNS13}). 

\subsection{H 1743$-$322}\label{s:src3}

\begin{figure}
\centerline{\includegraphics[width=6cm]{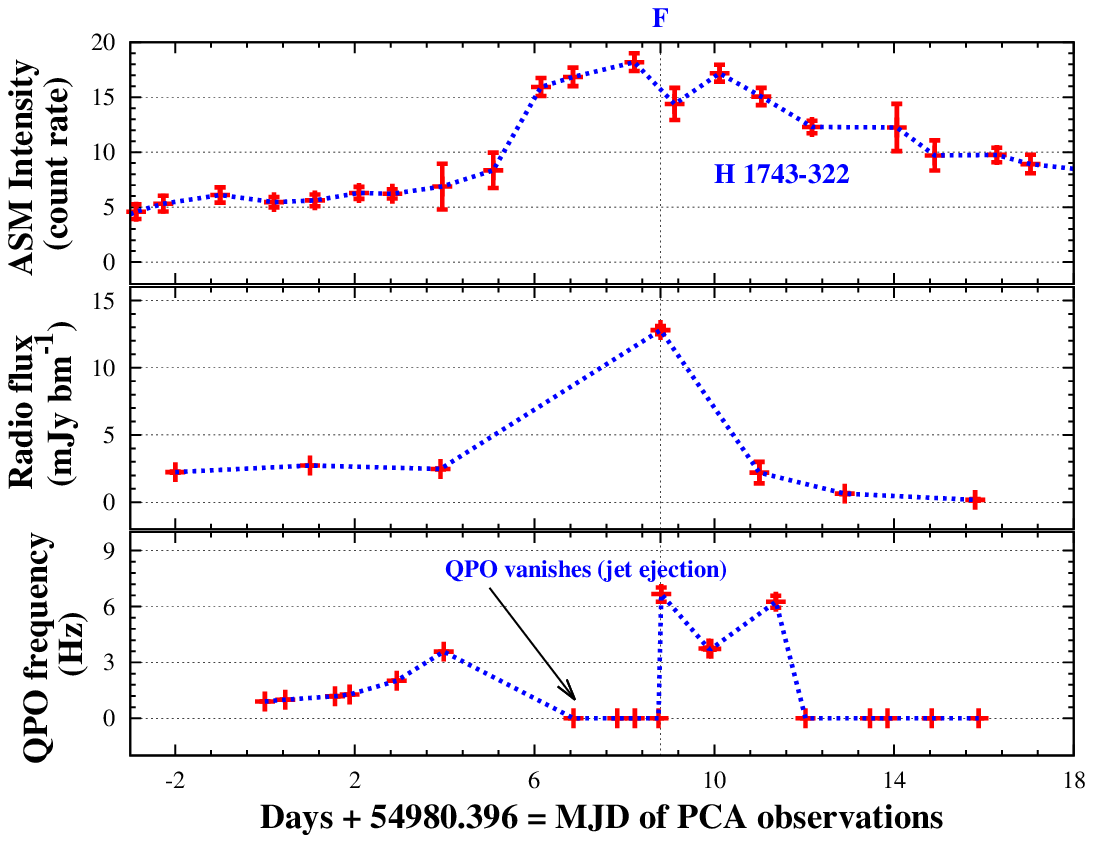} \qquad
            \includegraphics[width=6cm]{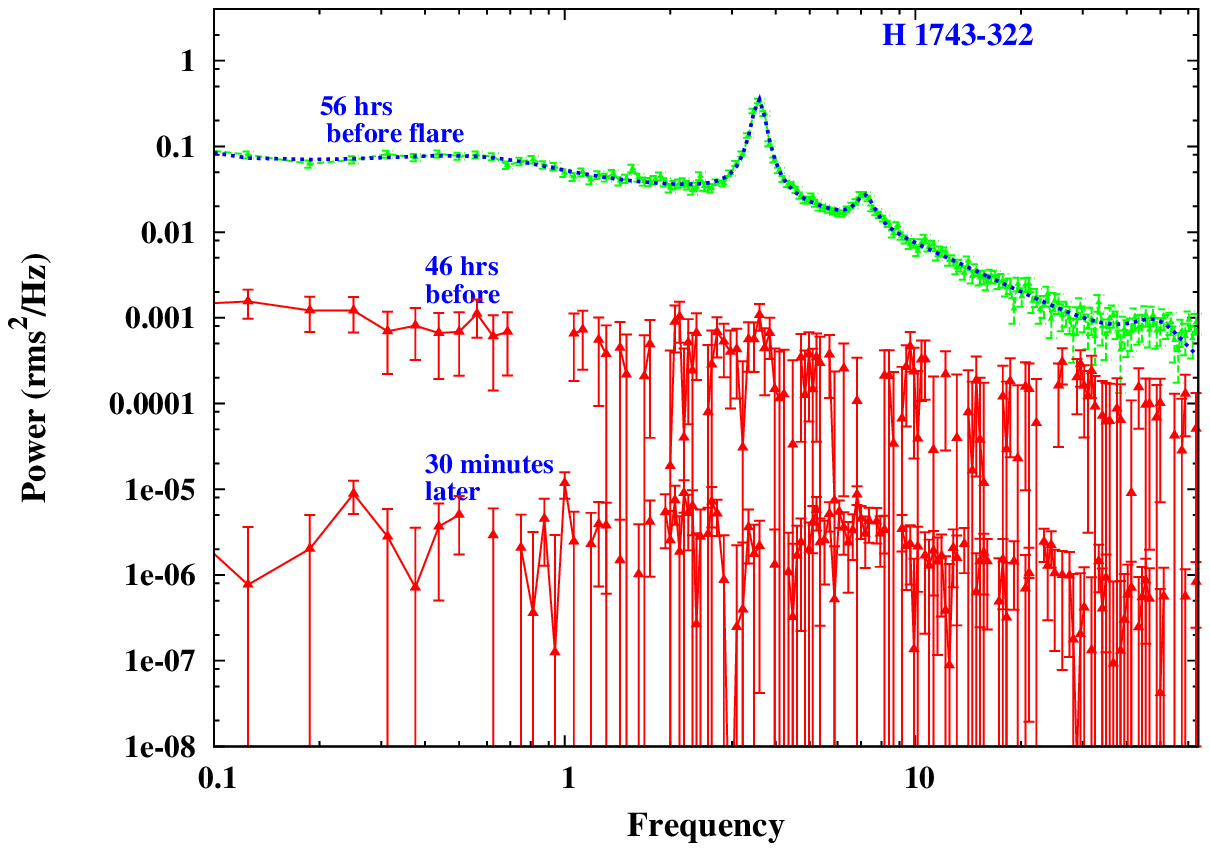}}
\caption{On the left side, we show variation of X-ray intensity, Radio flux and evolution of QPO 
frequencies, observed for the BH source H 1743$-$322 (2009 outburst). The evolution of the PDS (right side 
of the figure) showing the absence of QPOs during the flare.\label{f:3}}
\end{figure}

During the 2009 outburst of H 1743$-$322, we observed that the QPO frequency increases from 
0.91 Hz (MJD 54980.39) to 3.58 Hz (MJD 54984.37). The next observation (on MJD 54987.26) showed 
absence of QPOs (see right side of Figure \ref{f:3}) and spectral softening was observed around 
46 hrs before the peak Radio flare (primary ejection) of 12.8 mJy bm$^{-1}$ at 8.4 GHz 
(see also \citealt{MJ12}). We also noted that before the flare, the QPO observed was of type C, 
whereas after the flare QPO was of type B. Complete absence of QPO and spectral 
softening was noted after 3 days of QPO re-appearance. Spectral analysis also
suggest that the primary ejection could have triggered during the transition from hard to 
soft-intermediate state of 2009 outburst \citep{RD2013}. 

\subsection{GRO J1655$-$40}\label{s:src4}

\begin{figure}
\centerline{\includegraphics[width=6cm]{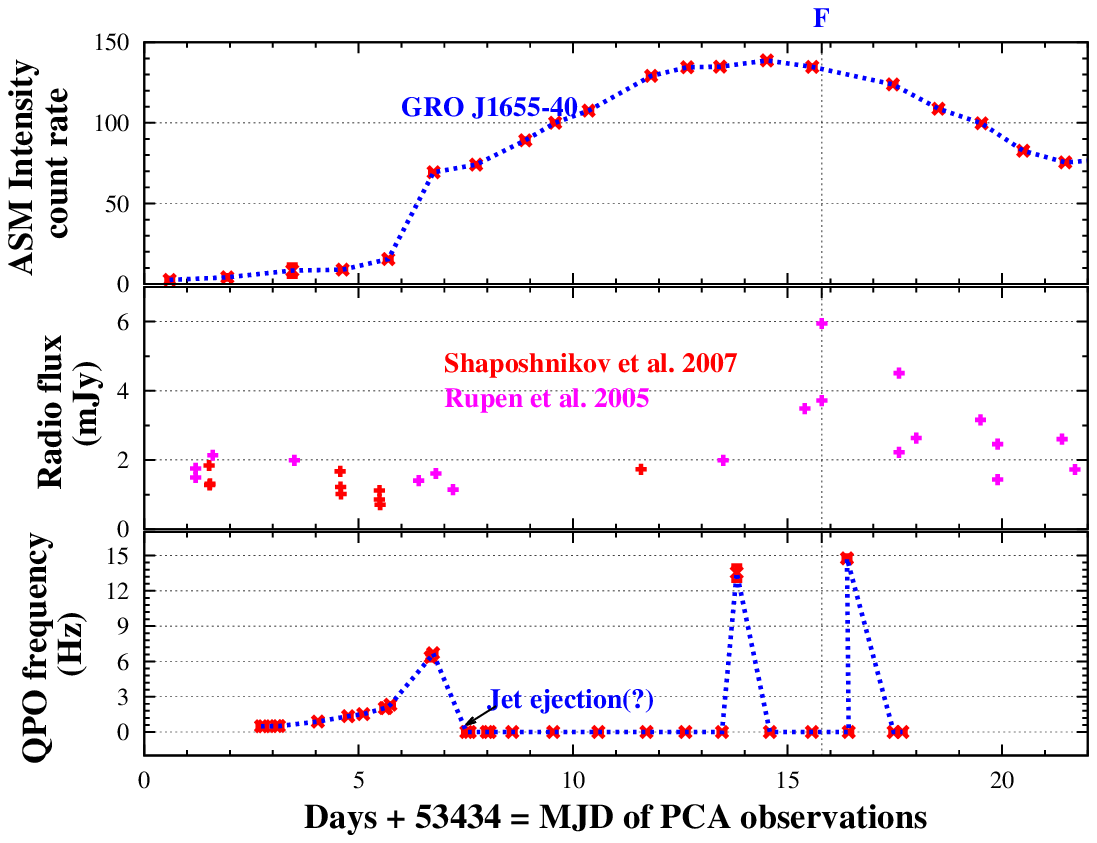} \qquad
            \includegraphics[width=6cm]{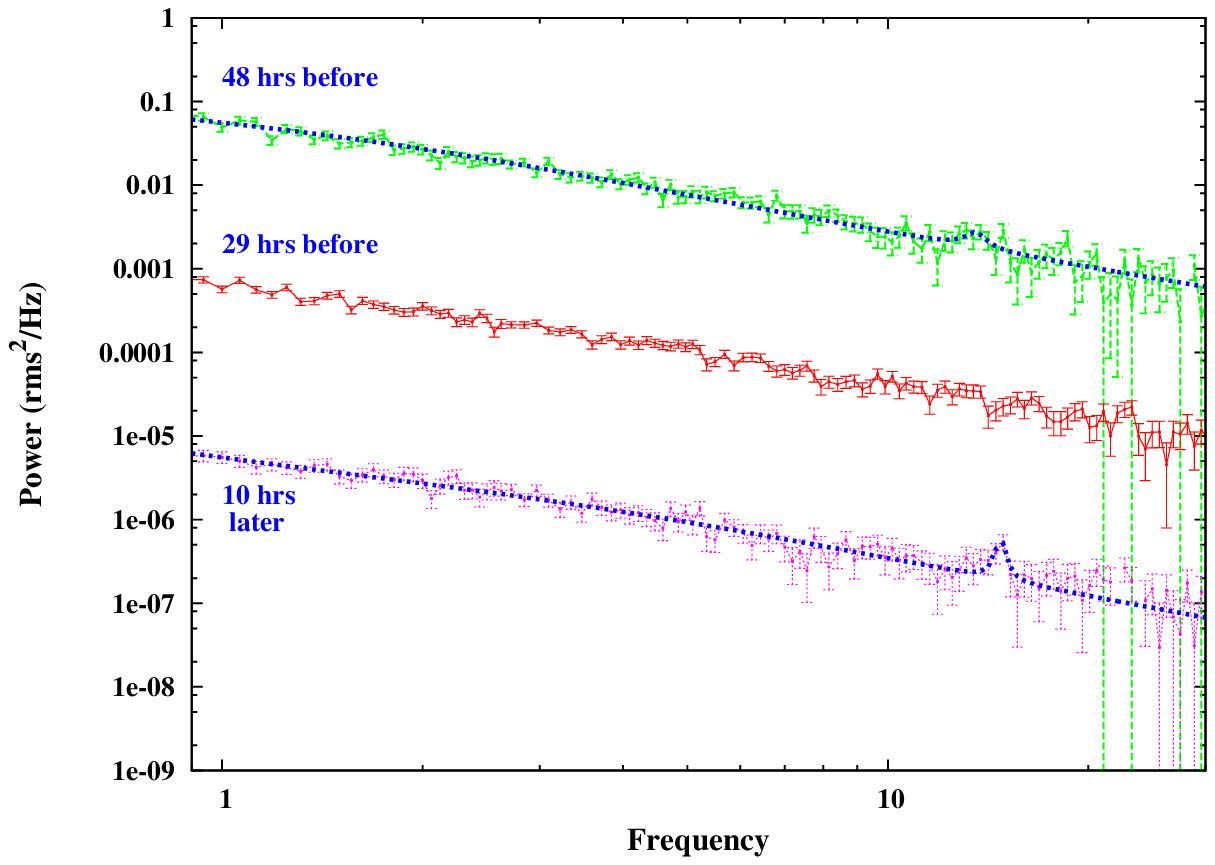}}

\caption{The variation of X-ray intensity, Radio flux and QPO frequencies observed for the BH source 
GRO J1655$-$40 (2005 outburst) is shown on the left side, while the evolution of power spectra during 
the peak flare is shown on the right side. \label{f:4}}
\end{figure}

During the 2005 outburst of GRO J1655$-$40, we observed an evolution of QPO frequency from 1 to 
6 Hz (see bottom panel on left side of Figure \ref{f:4}), followed by an absence of QPOs for almost 6 days. 
During the observation where the QPO was not present in the PDS, the ratio of soft to hard flux increased
and the outburst evolved from hard to soft-intermediate state. But since there is no Radio 
observation available, the time of possible ejection (as a radio flare) is not known. Around 6 days 
later (after the 1st X-ray peak of the 2005 outburst), we observed QPOs of 14 Hz in the power 
spectra around 48 hrs before a peak flare of 6 mJy (see middle panel of
left side figure). The observation 29 hrs before the flare showed absence of QPO. QPO re-appeared 
around 10 hrs after the flare (power spectral evolution is shown in right side of the figure). During this 
phase, we observed a decrease in total rms of PDS, absence of phase lag and increase in soft 
flux (3 - 20 keV) over hard X-ray flux (20 - 150 keV). The QPO observed before and after the ejection 
was possibly of type C*/C (high Q-factor with less rms).
    
\subsection{Other BH sources}\label{s:others}

Absence of QPOs during a flare event (i.e., transient Jets) has been already observed for the BH source 
GRS 1915$+$105 \citep{VS01,FM01}. We have extended our study to few more BH sources (eg. XTE J1752$-$223, 
XTE J1550$-$564, GX 339$-$4, MAXI J1836$-$194). Temporal analysis of these sources showed that 
whenever a Radio flare occurs, QPOs `disappear' along with reduction of total rms 
(see also \citealt{FP09}) of the power spectra. Absence of phase lag for soft to hard photons was 
also observed during the flares. Spectral analysis showed that there was an increase 
in thermal flux whenever a QPO was absent during the radio flare.
Detailed spectro-temporal analysis (during radio flares) of all the sources will be presented elsewhere 
\citep{RD2013}.   

\section{Possible Physical Scenario}\label{s:pps}

Several attempts have been made until now to understand the phenomenon of outflows \citep{KC13} and 
disk-jet coupling in BH 
binaries \citep{Meier2004,FP09}. But none of these studies have discussed 
the disk-jet coupling in the context of QPO `disappearance' and subsequent spectral softening. 
We attempted to understand the disk-jet connection \citep{VS01, AN01} in the outbursting sources based 
on the Two Component Advective Flow (TCAF) in the presence of magnetic field. According to TCAF 
\citep{CT95}, there are two types of flow viz, Keplerian and sub-Keplerian. The sub-Keplerian 
halo matter 
forms the Compton cloud (i.e., CENtrifugal pressure supported BOundary Layer (CENBOL)) during the
shocked-accretion phase. Due to oscillation of the shock, the CENBOL (i.e., also the store house of `hot' 
electron source) surface may oscillate resulting in QPOs.

A sudden occurrence of a radio flare can occur in the presence of magnetic field in the disk. According to 
the magnetised-TCAF \citep{AN01}, matter can anchor large stochastic magnetic fields during the phase
of accretion of matter from the companion. When the flux tube enters into the `hot' Compton cloud
of temperature $\sim 10^9 K$, it collapses catastrophically due to magnetic tension (i.e.,
the strongest force in the `hot' plasma within the CENBOL)
and evacuates the inner part of the disk producing Jets. As a result, CENBOL gets disrupted and hence 
its oscillation ceases, resulting in absence of QPO. As matter gets evacuated from the CENBOL in the 
form of Jets, the energy spectra will be dominated by thermal emission.

\begin{figure}
\centerline{\includegraphics[width=08cm, height=5cm, angle=0]{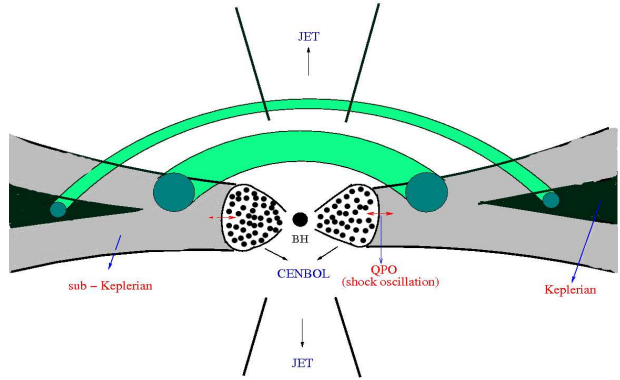}}
\caption{Magnetised-TCAF model, showing Keplerian and sub-Keplerian components of flow along with
magnetic flux tubes, which are responsible for disrupting the inner-part of the disk 
(i.e., CENBOL \citep{CT95}).
Adopted from \citealt{RN13}.}
\end{figure}

\section{Conclusion and Discussion}\label{s:conclu}

In order to verify the conjecture of `disappearance' of QPO during the Jet ejections (i.e., radio flares),
we investigated the X-ray properties of several BH sources. Our analysis seems to suggest that 
there is an absence of QPO, absence of soft/hard lags, decrease in total rms of the
power spectra, when an ejection occurs. The spectral evolution also implies the softening of the spectra 
during the ejections. Based on the magnetized-TCAF model, we 
can understand that the absence of QPO and spectral softening implies the disruption of inner-part of the 
disk in the presence of magnetic field. The re-appearance time scale of QPOs ($\sim$ hour to day) implies 
the time taken by the sub-Keplerian matter to form inner-part of the disk (i.e., the CENBOL).


\section*{Acknowledgements}

Authors are thankful to Dr. P. Sreekumar of ISRO Satellite Centre for support related to 
participation in this conference. This research has made use of the General High-energy Aperiodic 
Timing Software (GHATS) package developed by Dr. Tomaso Belloni at 
INAF - Osservatorio Astronomico di Brera.


\appendix


\label{lastpage}
\end{document}